\documentclass[11pt]{article}
\usepackage{authblk}

\usepackage{amsmath,graphicx}
\usepackage[round,numbers, sort&compress]{natbib}
\DeclareMathOperator*{\argmin}{\arg\!\min}

\makeatletter
\def\tagform@#1{\maketag@@@{[#1]\@@italiccorr}}
\makeatother
\usepackage{setspace}
\usepackage[margin=1in]{geometry}
\usepackage{amssymb}
\usepackage[outdir=./]{epstopdf}
\usepackage{subfigure}
\usepackage{tikz}
\usepackage{xcolor}
\usepackage[percent]{overpic}
\usepackage{amsthm}

\usepackage{algorithm}
\usepackage{algorithmic}
\let\Algorithm\algorithm
\renewcommand\algorithm[1][]{\Algorithm[#1]\setstretch{1.4}}
\title{\uppercase{ \bf \large \center SMS MUSSELS: A Navigator-free Reconstruction for Simultaneous MultiSlice Accelerated MultiShot Diffusion Weighted Imaging }}

\author[1]{\it \small Merry Mani}
\author[2]{\it \small Mathews Jacob}
\author[3]{\it \small Graeme McKinnon}
\author[3]{\it \small Baolian Yang}
\author[4]{\it \small Brian Rutt}
\author[4]{\it \small Adam Kerr}
\author[1]{\it \small Vincent Magnotta}
\affil[1]{\it \small Department of Radiology, University of Iowa, Iowa City, Iowa}
\affil[2]{\it \small Department of Electrical and Computer Engineering, University of Iowa, Iowa City, Iowa}
\affil[3]{\it \small Global Applied Science Laboratory, GE Healthcare}
\affil[4]{\it \small Department of Radiology, Stanford University}

\begin{document}
\date{}
\maketitle

\vspace{30mm}
\noindent  Correspondence to :\\
Merry Mani\\
L309 Papajohn Biomedical Discovery Building,\\
169 Newton Road\\
Iowa City, Iowa, 52242\\

\noindent  email: merry-mani@uiowa.edu \\
phone number: (319) 335-9569.\\
\\
Word count : 4980\\
figures+ tables count : 9\\
\\
Running title: Joint reconstruction for SMS-accelerated MS-DWI using annihilating filter k-space formulation \\

\newpage
\noindent { \bf Abstract}

\noindent {\bf Purpose:} To introduce a novel reconstruction method for simultaneous multi-slice (SMS) accelerated multi-shot diffusion weighted imaging (ms-DWI). \\

\vspace{-1.5em}\noindent {\bf Methods:} SMS acceleration using blipped CAIPI schemes have been proposed to speed up the acquisition of ms-DWIs. The reconstruction of the above data requires (i) phase compensation to combine data from different shots and (ii) slice unfolding to separate the data of different slices. The traditional approach is to first estimate the phase maps corresponding to each shot from a navigator or from the slice-aliased individual shot data. The phase maps are subsequently fed to a iterative reconstruction scheme to recover the slice unfolded DWIs. We propose a novel reconstruction method to jointly recover the slice-unfolded k-space data of the multiple shots. The proposed reconstruction is enabled by the low-rank property inherent in the k-space samples of a ms-DW acquisition. Specifically, we recover the missing samples of the multi-shot acquisition by enforcing a low-rank penalty on the block-Hankel matrix formed by the k-space data of each shot for each slice. The joint recovery of the slice-unfolded k-space data is then performed using a SENSE-based slice-unfolding subject to the low-rank constraint. The proposed joint recovery scheme is tested on simulated and in-vivo data and compared to similar un-navigated methods at slice acceleration factors of 2 and 3.\\

\vspace{-1.5em}\noindent {\bf Results:} Our experiments show effective slice unaliasing and successful recovery of DWIs with minimal phase artifacts using the proposed method.  The performance is comparable to existing methods at low acceleration factors and better than existing methods as the acceleration factor increase.\\

\vspace{-1.5em}\noindent {\bf Conclusion: } For the slice acceleration factors considered in this study, the proposed method can successfully recover DWIs from SMS-accelerated ms-DWI acquisitions.\\

\vspace{2em}

\newpage

\noindent {\Large Introduction}\\ 

\vspace{-1.5em}\noindent Diffusion-weighted imaging (DWI) is widely employed in clinical and research studies to assess brain white matter microstructure. Conventional diffusion weighted (DW) data collection is performed using single-shot EPI (ss-EPI) acquisition primarily due to its immunity to bulk motion. However, the low bandwidth of ss-EPI along the phase encoding direction leads to geometric distortions stemming from B0 field inhomogeneity. Similarly, the T2* decay accompanying the long EPI readout results in blurring in the images.  To reduce the severity of geometric distortions and T2* related blurring, ss-EPI trajectories are coupled with partial Fourier imaging and other parallel imaging based in-plane acceleration methods such that short read-out times can be achieved.  However, for high spatial resolution applications, this calls for high acceleration factors that results in significant signal-to-ratio (SNR) loss. Furthermore, the SNR benefits of the ultra-high field scanners also cannot be fully harnessed to achieve spatial resolution using ss-EPI techniques; the amplified T2* decay induced blurring and B0 induced geometric distortion significantly degrade the image quality at low acceleration factors \cite{Heidemann2012,Jeong2013,Wu2017}.  These factors have constrained the imaging resolution of DWIs to approximately 2~mm isotropic, which impose significant limitations on diffusion based microstructure studies where partial volume effects obscure the characterization of tissue microstructure in the brain. \\

\vspace{-1.5em}\noindent Multi-shot EPI (ms-EPI) based diffusion weighted imaging (ms-DWI), on the other hand, can overcome several of the limitations of ss-EPI acquisitions \cite{Wu2017}. The segmented read-outs increase the bandwidth along the phase encoding direction. Without worsening the EPI-related artifacts, ms-EPI can thus enable higher spatial resolution diffusion imaging on low and high-field strengths. Moreover, the shorter echo-time (TE) limits the blurring and the multiple shots buys back the SNR, both of which are crucial for high resolution imaging. However, this comes at the expense of long scan times since the repetition times (TRs) are increased by a factor proportional to the number of shots compared to ss-EPI schemes. This low SNR efficiency limits the ability of ms-DWI to enable advanced diffusion protocols with high angular resolution in reasonable scan time. Replacing the conventional 2D slice-by-slice imaging with simultaneous multi-slice (SMS) imaging \cite{Larkman2001a,Breuer2005a,Setsompop2012,Feinberg2013,Barth2016} can significantly accelerate ms-DWI imaging while improving the SNR efficiency \cite{Frost2015,Dai2017,Hashemizadehkolowri2018,Herbst2017,Chang2015}.\\

\vspace{-1.5em}\noindent Multi-shot diffusion weighted data acquired in slice-by-slice or SMS mode require specialized reconstructions involving phase calibration to compensate for the phase differences between the data of each shot. This is because the k-space data of each of the shots are encoded in separate TRs, leading to a unique phase accrual for the data from each of the k-space shots. Due to this phase mismatch, the k-space data from the separate shots cannot be directly combined together. One simple solution is to reconstruct each of the shots separately assuming an accelerated acquisition either using SENSE \cite{Pruessmann1999} or GRAPPA \cite{Griswold2002}-based methods and then combine the resulting magnitude images. This technique can also be extended to reconstruct SMS-accelerated MS-DW acquisition that also incorporates slice unfolding \cite{Zhu2012,Setsompop2012,Blaimer2006,Cauley2014}. Since the phase information is completely discarded, the reconstructed magnitude-combined images will be devoid of artifacts arising from motion-induced phase errors. However, the high in-plane acceleration factors involved in such reconstructions will result in unacceptable geometry-factors (g-factors).  \\

\vspace{-1.5em}\noindent To overcome the problems associated with high g-factors, reconstruction involving data from all shots are desired. Two classes of methods have been proposed for multi-shot data reconstruction incorporating data from all shots : (i) multi-stage methods involving explicit phase compensation \cite{DeCrespigny1995,Ulug1995,Butts1996, Jeong2013,Liu2004,Nunes2005,Pipe2002, Uecker2009a, Chen2013,Chu2015,Guhaniyogi2015,Chang2015,Holdsworth2009} and (ii) joint reconstruction methods not requiring phase calibration \cite{Mani2017a, Mani2016b,Hu2018}. Most reconstruction methods that have been developed so far for SMS-accelerated ms-DWI reconstruction fall in the first category \cite{Frost2015,Dai2017,Hashemizadehkolowri2018,Herbst2017,Chang2015}. These methods undertake a two-stage approach where in the first stage, the phase associated with each shot is calibrated. In the second stage, the phase correction and slice unfolding are performed simultaneously in a generalized iterative GRAPPA based \cite{Frost2015,Dai2017,Hashemizadehkolowri2018} or SENSE based \cite{Herbst2017,Chang2015} reconstruction using the calibrated phase maps. While such reconstructions provide better SNR than the individual shot reconstructions, the accuracy of the phase estimates determines the quality of the final reconstruction. Phase calibration can be done using navigator based methods or navigator-free (self-navigated) methods and each method has its own advantages and disadvantages. Phase calibration using navigator based approaches require additional data acquisition to capture the shot-to-shot phase variations. While these methods can support high acceleration factors \cite{Dai2018}, navigators increases the effective TR, increase the specific absorption rate (SAR) and may not accurately capture phase variations in the presence of bulk motion \cite{Wu2017}. Navigator-free phase calibration using parallel imaging is time efficient. However, such methods cannot support high acceleration factors since the phase estimates becomes highly erroneous as the number of shots and multi-band factor increase. In the latter class, a joint reconstruction of k-space data from all shots are accomplished exploiting data priors. Phase information is not required for reconstructing MS-DW data using this approach. For example, the recently proposed MUSSELS algorithm \cite{Mani2017a, Mani2016b} exploits the low-rank property that exists between the k-space data of individual shots to formulate the joint reconstruction of the multi-shot data as a structured low-rank matrix recovery problem. Similarly, the formulation in \cite{Hu2018} employs a locally low-rank prior. Such reconstructions are relatively immune to inaccuracies in phase errors and can provide robust reconstructions compared to navigator-free phase calibrated methods at high shot factors. \\

\vspace{-1.5em}\noindent  The aim of this work is to develop a joint reconstruction method for SMS-accelerated multi-shot DW data. In a single-band MS-DW acquisition, the phase varies from shot-to shot. This phase variation is different for each slice since the cardiac pulsation and respiratory effects vary for different slice locations. Hence, in an SMS-accelerated acquisition, the shot-to-shot phase variation for the different slices needs to be accounted for. We generalize the phase-calibration-free MUSSELS reconstruction for the case of SMS accelerated ms-DWI data. Specifically, we construct a block Hankel matrix using the k-space shot data for each slice. Then, a low-rank prior is enforced on the fully known block-Hankel matrix for each slice. The new joint reconstruction scheme for SMS accelerated ms-DWI thus augments a SENSE-based SMS slice unfolding with the low-rank prior for each slice. We show that the above joint reconstruction can effectively remove  aliasing artifacts arising from slice folding and motion-induced phase artifacts. We test the proposed reconstruction on simulations and in-vivo data. We compare the proposed reconstruction to the MUSE-based reconstruction for SMS-accelerated ms-DWI and show superior results. \\

\vspace{-0.5em}\noindent{ \Large Theory}\\

\vspace{-1.5em}\noindent{  \bf  SMS accelerated ms-DWI Acquisition using blipped CAIPI }\\

\vspace{-1.5em}\noindent Single shot EPI sequences have been accelerated for simultaneous multi-slice imaging using blipped CAIPI techniques. Here a spatial shift in the phase encoding direction between simultaneously excited slices is achieved by modulating phase of the k-space data of the slices by a phase factor, as per the Fourier shift theorem. The phase factor is chosen to improve the g-factor of the unfolded slice images. Accordingly, the standard approaches shift the simultaneously excited slices by a factor of FOV/L for the case of $L$ slices. When $L$ is low, this approach provides a reasonable improvement in the g-factor.
%In general, to improve the g-factor, an FOV shift of FOV/L is chosen for the case of $L$ slices. 
Then, the measured k-space data for the $n^{th}$ phase encoding line can be written as 
\begin{equation}
\label{equation 1}
s(k_x,n\Delta k_y) = \sum_{l=1}^L~s_l(k_x,n\Delta k_y).\text{exp}({-j n\Delta k_y.\underbrace{(l-1)(2\pi/L)}_{\theta{_l}(k_x,n\Delta k_y)}}), \\
\end{equation}
where $s_l$ is the signal from the $l^{th}$ slice, $s$ is the sum of the signals measured from $L$ slices, ${\theta{_l}(k_x,n\Delta k_y)}$ is the phase factor and $k_x$ and $n\Delta k_y$ are the k-space indices. In practice, this is achieved in the  blipped CAIPI techniques by adding a phase cycled train of extra gradient blips along the slice selection gradient $G_z$ simultaneously with the phase encoding blips in the $G_y$ axis of the EPI readout. The $G_z$ gradient moments are designed to create the desired phase accrual required to shift the slices along the phase encoding direction. \\

\vspace{-1.5em}\noindent For ms-EPI acquisition, the above equation is slightly modified. This is because, for shot $i$, the phase encoding index $n$ changes to
 \begin{equation}
n' = (n-1)N_s +i,
\end{equation}
leading to re-write Eq. \ref{equation 1} as:
\begin{equation}
\label{equation 2}
s_i(k_x,n'\Delta k_y) = \sum_{l=1}^L~s_{l,i}(k_x,((n-1)N_s+i)\Delta k_y).\text{exp}({-j ((n-1)N_s+i)\Delta k_y(l-1)(2\pi/L)}),
\end{equation}
where $N_s$ is the number of shots. Hence, the shifting in blipped CAIPI method needs to be modified for ms-EPI acquisition to achieve a shift of FOV/L for the alias-free image by modifying the $G_z$ gradients.   The specific modifications in the pulse sequence used to acquire ms-DW data using blipped CAIPI are detailed in \cite{Dai2017}. Specifically, the gradient blip amplitude has been multiplied by the multi-shot factor and the phase offset for each shot has been modified based on the number of shots.\\

\vspace{-1.5em}\noindent{  \bf SMS accelerated ms-DWI recovery }\\

\vspace{-1.5em}\noindent In the following section, we examine the phase relations that exist between the data generated from a SMS accelerated ms-DW acquisition. We assume a multi-band RF pulse that simultaneously excited $L$ slices using the modified blipped CAIPI technique.  Then, a given k-space readout will sample the data from the $L$ simultaneously excited slices. For an $N_s$ shot acquisition, 
the k-space readout is split into $N_s$ segments and are sampled over $N_s$ separate TRs.
 Let the {\it magnitude} diffusion weighted image of a given slice $l$ and a given diffusion direction be denoted as $\rho_l({\bf x})$, where $\bf{x}$ represents spatial co-ordinates. The slice-unfolded {\it complex} diffusion weighted image corresponding to the $i^{th}$ k-space shot from slice $l$ can be written as 
\begin{equation}
	\label{first_eq1}
	 m_{l,i}({\bf x})=\rho_l({\bf x})~\phi_{l,i}({\bf x}); \forall {\bf x}, ~~~~~ l=1: L,
	 \end{equation}
where $\phi_{l,i}({\bf x})$ is the phase associated with the DWI of slice $l$ and shot $i$. Notably, under the influence of the strong diffusion encoding gradient, a subsequent k-space readout of the same data $\rho_l({\bf x})$, (from the same slice and diffusion gradient) will have a different phase associated with it due to the inter-shot motion of the spins that are being encoded by the diffusion gradients \cite{Anderson1994,Liu2005,Eichner2015}.
Thus, the measured {\it complex }DWI from the $n^{th}$ shot for the above slice and diffusion encoding will have a different a phase $\phi_{l,n}({\bf x})$ associated with it , leading us to write

\begin{equation}
	\label{first_eq2}
	 m_{l,n}({\bf x})=\rho_l({\bf x})~\phi_{l,n}({\bf x}); \forall {\bf x}, ~~~~~ l=1: L.
	 \end{equation}
In the special case where SMS-accelerated acquisition is (nearly) fully sampled (e.g., a single shot acquisition), the magnitude DWI $\rho_l({\bf x})$ corresponding to each slice can be obtained by simply ignoring the phase of the slice unfolded complex DWIs $m_{l}({\bf x})$.
 However, in the case of a multi-shot acquisition, the complex image $m_{l,n}({\bf x})$ corresponding to each shot needs to be reconstructed from the under-sampled individual shot-data first, before the inter-shot phase can be eliminated.\\

\vspace{-1.0em}\noindent{  \bf Generalization of MUSSELS formalism for SMS data}\\

\vspace{-1.5em}\noindent It is important to note that although there is a unique phase accrual for the data acquired with each shot from a given slice, the magnitude of the DWI acquired using each shot is the same. Thus, significant redundancy exists between the samples acquired using the different k-space shots. In our previously introduced MUSSELS method \cite{Mani2017a}, we formalized this redundancy as a low-rank property of the measured k-space data. This property is easily generalizable to a SMS framework.  Specifically, since there is no interaction between the phase of the data from different slices, annihilation relations of the form :
	\begin{equation}
	\label{first_eq}
	  m_{l,i}({\bf x})\phi_{l,n}({\bf x})-m_{l,n}({\bf x})~\phi_{l,i}({\bf x}) = 0; \forall \bf x.
	 \end{equation}
can be established between DWI of every pair of shot images for each slice, $l=1: L$. Taking the Fourier transform on both sides of \eqref{first_eq}, we obtain annihilation relations between the k-space data of each shots:
	\begin{equation}
	\label{first_eq3}
	 \widehat{m_{l,i}}[{\bf k}]*\widehat{\phi_{l,n}}[{\bf k}]-\widehat{m_{l,n}}[{\bf k}]*\widehat{\phi_{l,i}}[{\bf k}] = 0; \forall \bf k,
	 \end{equation}
	 where $\widehat{m_{}}[\bf k]$ and $\widehat{\phi_{}}[\bf k]$ denote the Fourier coefficients of ${m_{}}(\bf x)$ and ${\phi_{}}(\bf x)$, respectively. To formalize the low-rankedness between the acquired k-space data of each shots, we re-write the above convolution relation as a multiplications involving Hankel matrices:
	 \begin{equation}
	\label{sec_eq}
	{\mathbf{H}}(\widehat{{m}_{l,i}})\cdot \widehat{\boldsymbol\phi}_{l,n}-{\mathbf {H}}(\widehat{{m}_{l,n}})\cdot \widehat{\boldsymbol\phi}_{l,i} = \bf 0. 
	\end{equation}
where ${\bf{H}}(\widehat{m_{l,i}})$ is a block Hankel matrix of size $(N_1-r+1)(N_2-r+1)\times r^2 $ generated from the $N_1 \times N_2$ Fourier samples of $\widehat {m_{l,i}}[\bf{k}]$; r being the support of the $\widehat{\phi_{l}}[\bf{k}]$. Since, this relation holds true for all pairs of shots for any given slice $l$, we can combine the annihilation relations in the matrix form as:\\
\begin{equation}
	\label{equation 3}
	\underbrace{\begin{bmatrix} {\bf{H}}(\widehat{m_{l,1}}) & {\bf{H}}(\widehat{m_{l,2}}) & ... & {\bf{H}}(\widehat{m_{l,N_s}}) \end{bmatrix}}_{{{\bf{H}}}_{l}(\bf\widehat{m})}\underbrace{\left[ \begin{array}{c} \widehat{\boldsymbol\phi_{l,2}} \\ -\widehat{\boldsymbol\phi_{l,1}}\\ 0 \\ 0\\  \vdots \\ 0 \end{array} \begin{array}{c} 0\\
 \widehat{\boldsymbol\phi_{l,3}} \\ -\widehat{\boldsymbol\phi_{l,2}}\\ 0 \\ \vdots \\ 0 \\ \end{array}\begin{array}{c} \cdots \end{array}
\begin{array}{c} 0\\0 \\ \vdots \\ 0 \\ \widehat {\boldsymbol\phi_{{l,N_s}}}\\
-\widehat {\boldsymbol\phi_{{l,N_s-1}}}  \end{array}\begin{array}{c}  \widehat{\boldsymbol\phi_{l,3}} \\0\\-\widehat{\boldsymbol\phi_{l,1}} \\ 0\\  \vdots \\ 0 \end{array}  \begin{array}{c} \cdots \end{array} \right]}_{\bf{\hat{\pmb \phi_l}}}= \begin{bmatrix}0 & 0 & ...& 0 & 0 ... \end{bmatrix}. \\
\end{equation}
The above relation, compactly expressed as ${{\bf H}}_l({\bf \widehat{m}})~\hat{{\pmb \phi_l}} = \bf 0$, formalizes the redundancy in the measured k-space data of each shots as the low-rankedness of the structured matrix ${{\bf{H}}}_l(\bf \widehat{m})$. For the case of a multi-shot acquisition, the block-Hankel matrix ${{\bf{H}}}_l(\bf \widehat{m})$ will have a large number of missing entries since the k-space data of each shot $\widehat{{\bf m}_{l,i}}[\bf k]$ is under-sampled. However, the above relation requires that a fully filled block-Hankel matrix should satisfy the above low-rank property. Thus, the low-rankedness of the {\it fully known} block-Hankel matrix provides a powerful reconstruction constraint to recover missing k-space samples in each of the multi-shot k-space acquisitions, while preserving the SNR. Since the above criteria is true for the data of all slices, the low-rankedness of ${{\bf{H}}}_l({\bf \widehat{m}}), l=1:L$ provides an additional criteria to unfold the SMS-accelerated multi-shot data as a single-step recovery scheme that does not involve phase calibration. \\

\vspace{-1.5em}\noindent{  \bf SMS MUSSELS}\\

\begin{figure}
\includegraphics[trim = 0mm 0mm 0mm 0mm, clip, width=1\textwidth]{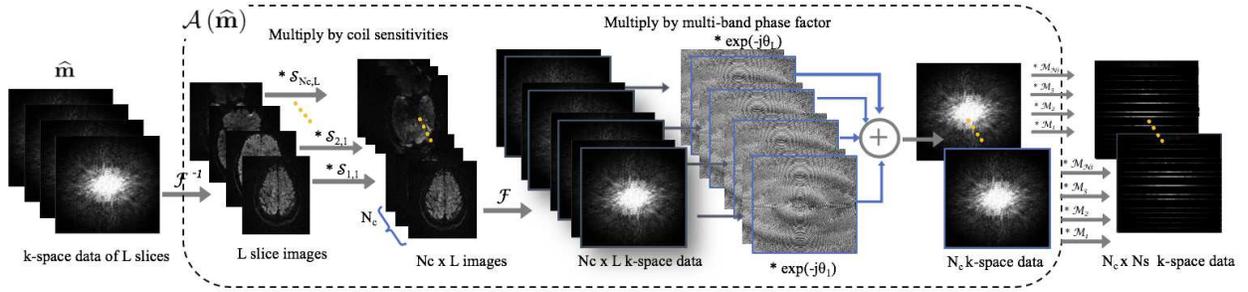}
\caption{Forward model for SMS MUSSELS involves taking (i) the inverse Fourier transform (${\cal {F}}^{-1}$) of the k-space data of L slices, (ii) multiplying the resulting image data by the coil sensitivities (${\cal {S}}$), (iii) taking the Fourier transform (${\cal {F}}$) to get the k-space channel-by-channel data of each slice, (iv) introducing slice shift by multiplying each slice data by phase factors ($\theta_l$) and adding the data from all slices in a channel-by-channel manner and finally (v) multiplying each channel data by a sampling mask (${\cal {M}}$) corresponding to the multi-shot acquisition.}
\label{fig:fig1}
\end{figure}

\vspace{-1.5em}\noindent We propose to augment the SMS slice unfolding with the low-rank constraint derived above to design a new joint reconstruction scheme for SMS-accelerated ms-DW data. We make use of the SENSE-based formalism explicitly making use of the coil sensitivity information. Specifically, we formulate the SMS-MUSSELS recovery as:

\begin{equation}
\label{use Hankel}
\widehat{\tilde {\bf m}}=  \argmin_{{\bf{\widehat{m}}}} \underbrace{||{\mathcal A}\left({\bf{\widehat{m}}}\right)-{\bf{\widehat{y}}}||^2_{\ell_2}}_{\mbox{data consistency}} + \lambda \underbrace{\sum_{l=1}^L||{\bf H}_l({\bf{\widehat{{m}}}})||_*}_{\mbox{low rank penalty}} .
\end{equation}
Here, $\widehat{{\bf m}}$ is the slice-unfolded k-space data of each shot. The operation ${\mathcal A}\left({\bf{\widehat{m}}}\right)$ incorporates SMS slice folding and enforces data consistency with the measured 
 SMS-accelerated multi-channel multi-shot k-space data, $\bf{\widehat y}$, of dimension $N_1\times N_2/N_s \times N_c\times N_{s}$. Here, $N_1\times N_2$ represents the size of the DWIs, $N_c$ is the number of channels and $N_s$ is the number of shots. The different steps in involved in the computation of ${\mathcal A}\left({\bf{\widehat{m}}}\right)$ are illustrated in figure \ref{fig:fig1}.  $ \lambda$  is a regularization parameter. ${{\bf{H}}}_l({\bf \widehat{m}})$ is the block Hankel matrix made up of the k-space data of each shot for slice $l$. The second term enforces low-rank property of the block-Hankel matrix for each slice. \\
 
\vspace{-1.5em}\noindent  The above cost function can be solved iteratively using alternating minimization schemes. Specifically, we adopted an augmented Lagrangian based scheme for minimizing the above cost function. The pseudo-code for the above algorithm can be found in \cite{Mani2017a}, with the operator ${\mathcal A}$ and ${{\bf{H}}}_l$ modified as described above.\\

\vspace{-0.5em}\noindent { \Large Methods}\\

\vspace{-1.5em}\noindent { \bf Experiments and datasets for validation}\\

\vspace{-1.5em}\noindent Blipped-CAIPI acquisition for multi-shot spin-echo EPI was implemented to study the utility of the technique for accelerating ms-DWI. Data were obtained from healthy volunteers in accordance with the Institutional Review Board of the University of Iowa. Imaging was performed using GE 3T MR750W scanner (GE Healthcare, Waukesha) with maximum gradient amplitude of 33 mT/m and slew rate of 120 T/m/sec employing a 32-channel receive coil.\\

\vspace{-1.5em}\noindent { \bf  In-vivo data}\\

\vspace{-1.5em}\noindent We tested the proposed SMS MUSSELS reconstruction on in-vivo data obtained in the following manner. The spin-echo diffusion weighted EPI scan was performed using a Stajeskal-Tanner sequence employing single-shot, 2-shot and 4-shot readouts in single-band and multi-band modes at MB = 2 and 3.  90$^\circ$ sinc pulses and 180$^\circ$ VERSE pulses \cite{Conolly1988} respectively were frequency modulated and summed to generate the multi-band RF pulses for the spin-echo excitation. Multi-band factors beyond three were not considered in this study due to SAR limitations. The simultaneously excited axial slices were 65mm apart at MB factor of 2 and 45mm apart at MB factor of 3. DW imaging parameters included: b-value=1000s/mm2; FOV = 220~mm~x~220~mm; matrix size=128~x~128; slice thickness~=~2~mm; and TE~=~57-142~ms depending on the number of shots, with a partial Fourier factor of .6875. All data consisted of twenty five diffusion weighted acquisitions and one non-diffusion weighted acquisition. The data acquisition time for the single-shot single-band data was 3.23 mins, 2-shot data at MB=1,2,3 were 5.24 mins, 3.09 mins and 2.10 mins respectively and 4-shot data at MB=1 and 3 was 9 mins and 3.58 minutes respectively.\\

\vspace{-1.5em}\noindent { \bf  Simulations}\\

\vspace{-1.5em}\noindent 
To study the utility of the proposed SMS MUSSELS reconstruction, we rely on simulations. Although one might think that a single-band multi-shot acquisition can act as a ground truth for comparing the performance of in-vivo multi-band experiments, this is challenging for several reasons. First of all, if there is subject motion between the single-band and multi-band acquisitions, the same slice encoding cannot not be achieved. Second, the inter-shot motion in these two acquisitions will be different. Hence the multi-shot phase error incurred in the two acquisitions will be different. For reasons, we make use of simulations to quantify errors in SMS-MUSSELS reconstruction. We will rely on single-band multi-shot acquisitions for simulating multi-band acquisitions. Data from different slices that are spatially apart were chosen to simulate a multi-band acquisitions as follows.  The channel-by-channel k-space data from each shot for the different slices were multiplied by a phase factor to spatially distribute the slice data and added along the slice dimension. Two- and four shot acquisitions were used to study reconstructions at multi-band factors 2 and 3. The simulated data were used in the SMS MUSSELS reconstruction to unfold the slices and recover the phase artifact free DWIs. The reconstructed images was compared to the ground truth images for quantification of accuracy using root-means-square error (RMSE) between the single-band DWIs and the reconstructed DWIs. The "ground truth images" are reconstructed using the 2D MUSSELS method. We also matched slice spacing of the simulations to the in-vivo experiments. \\

\vspace{-1.5em}\noindent { \bf  Reconstruction}\\

\vspace{-1.5em}\noindent The k-space data obtained from the multi-shot acquisitions were first corrected for Nyquist ghost artifacts resulting from odd-even shifts of the EPI acquisition using standard 1-D reference scan methods. The non-diffusion weighted (b0) images were reconstructed using a sum-of-squares (SOS) scheme. Coil sensitivity maps were estimated from the channel images of the non-diffusion weighted images.\\

\vspace{-1.5em}\noindent To compare the proposed reconstruction results, a second method proposed in Herbst et al., \cite{Herbst2017} was also implemented. Note that this method is different from the image domain MUSE implementation as proposed in \cite{Chang2015}. Notably, Chang et al., did not use a blipped-CAIPI implementation and extends the image domain MUSE to reconstruct the multiplexed EPI data. In contrast,  \cite{Herbst2017} use the composite sensitivity method originally proposed in \cite{Liu2005c} for spiral-based multi-shot diffusion weighted images, to reconstruct their blipped CAIPI data. For clarity, since \cite{Herbst2017} refers to this as generalized SMS MUSE, we will refer to this method as SMS MUSE.\\

\vspace{-1.5em}\noindent All reconstructions were implemented in MATLAB 2017a (The Mathworks, Natick, MA) on a desktop PC with an Intel i7-4770, 3.4 GHz CPU with 8 GB RAM. The SMS MUSSELS reconstruction is solved using an augmented Lagrangian scheme \cite{Bertsekas1976,Ramani2011}. A filter size of 4 x 4 was used for the 2-shot and and 6 x 6 was used for the 4-shot data. Although the low-rankedness can be imposed in a slice-based manner \cite{Park2017}, we chose to impose the same rank for every slice to reduce the manual tuning involved.  The reconstruction time varied from 2 mins to 8 mins per diffusion direction for various multi-shot and multi-band factors.\\ 

\vspace{-0.5em}\noindent{ \Large Results}\\

\vspace{-1.5em}\noindent { \bf  Simulations}\\

\begin{figure}
\includegraphics[trim = 0mm 0mm 0mm 0mm, clip, width=1\textwidth]{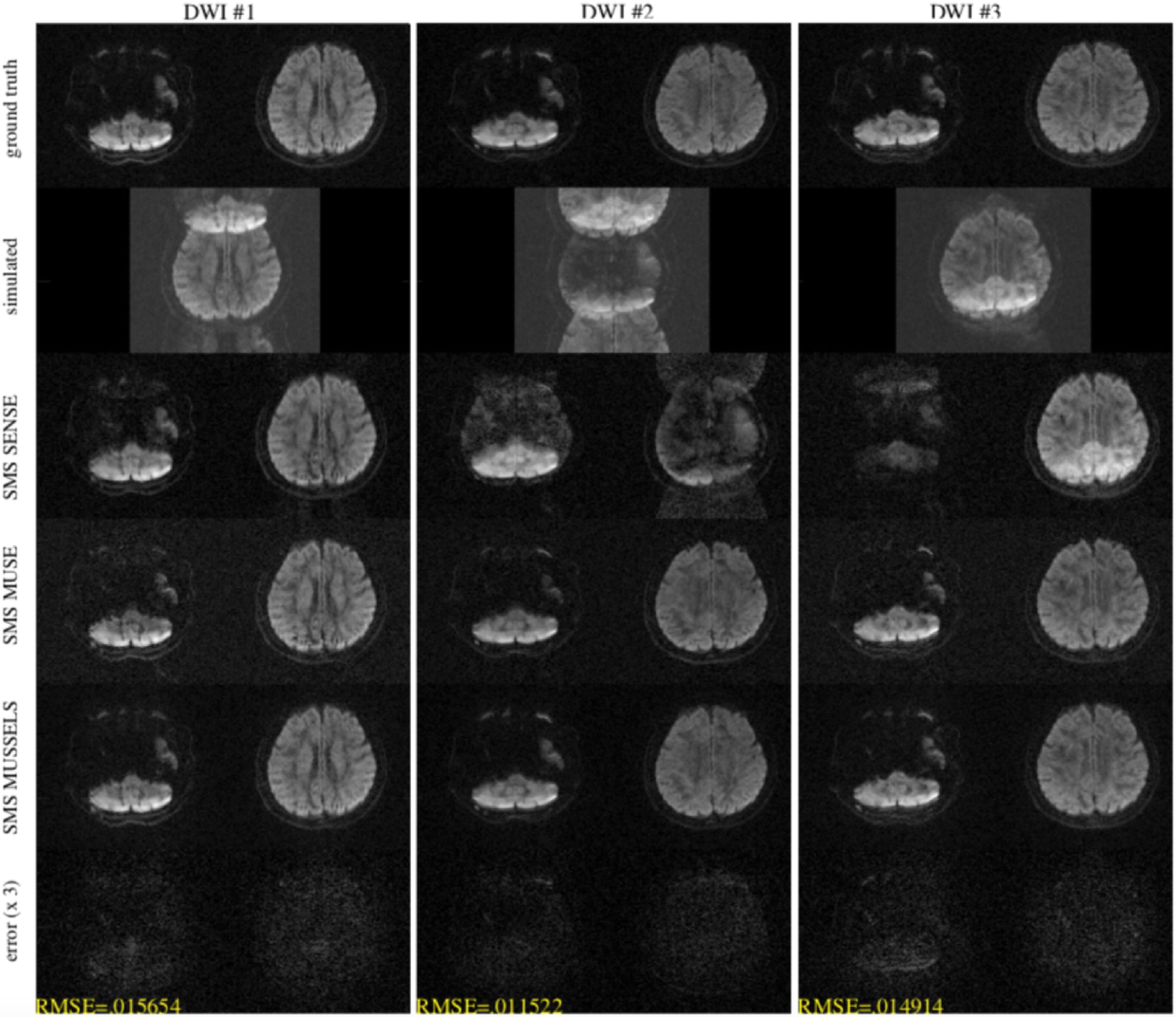}
\caption{Simulation for MB=2 from 2-shot DW data for three diffusion directions. The top row shows the two slices that were used for simulating the multi-band data. The second row shows the simulated data. The third row shows a SENSE reconstruction combing the k-space data from all shots without phase compensation. The fourth row shows the SMS MUSE reconstruction. The fifth row shows the SMS MUSSELS reconstruction and the last row shows the error in the SMS MUSSELS reconstruction compared to the ground truth.}
\label{fig:sim_ms2_mb2}
\end{figure}

\vspace{-1.5em}\noindent The experiments performed using 2-shot DW data to simulate a multi-band data at MB~=~2 are shown in figure 
\ref{fig:sim_ms2_mb2}. The top row shows the two slices that were used for simulating the multi-band data which we refer to as the ground truth data.   The multi-band data was simulated from the multi-channel 2-shot k-space data of the above slices. The second row displays the image corresponding to the simulated multi-band multi-shot acquisition, showing both slice aliasing and phase artifacts. Images corresponding to three different diffusion directions are shown as well  since the phase errors manifest differently depending on the diffusion encodings. In the third row, we show the results of a SMS SENSE reconstruction. Here, the k-space data from all shots are combined without phase correction and a SENSE-based slice unfolding is performed. If there are no phase errors between shots, we expect the SMS SENSE to provide reasonable reconstructions of the unfolded slices.
As evident from the figure, the SMS SENSE provide reasonable reconstruction for the  first diffusion direction, showing that the phase errors were not severe in this case. The SMS MUSE and the SMS MUSSELS reconstruction also provides reasonable results compared to the ground truth data. In the second and third diffusion direction (middle and third column), the SMS SENSE reconstruction show that there is significant inter-shot phase errors. The SMS MUSE and the SMS MUSSELS reconstruction were able to unfold the slices and recover the DWIs devoid of phase artifacts reasonably well. The difference image for the SMS MUSSELS reconstruction with respect to the ground truth data (magnified 3$\times$) also substantiate this. In all the cases, the RMSE is $\sim$ 1.5\%.\\

\begin{figure}
\includegraphics[trim = 0mm 0mm 0mm 0mm, clip, width=1\textwidth]{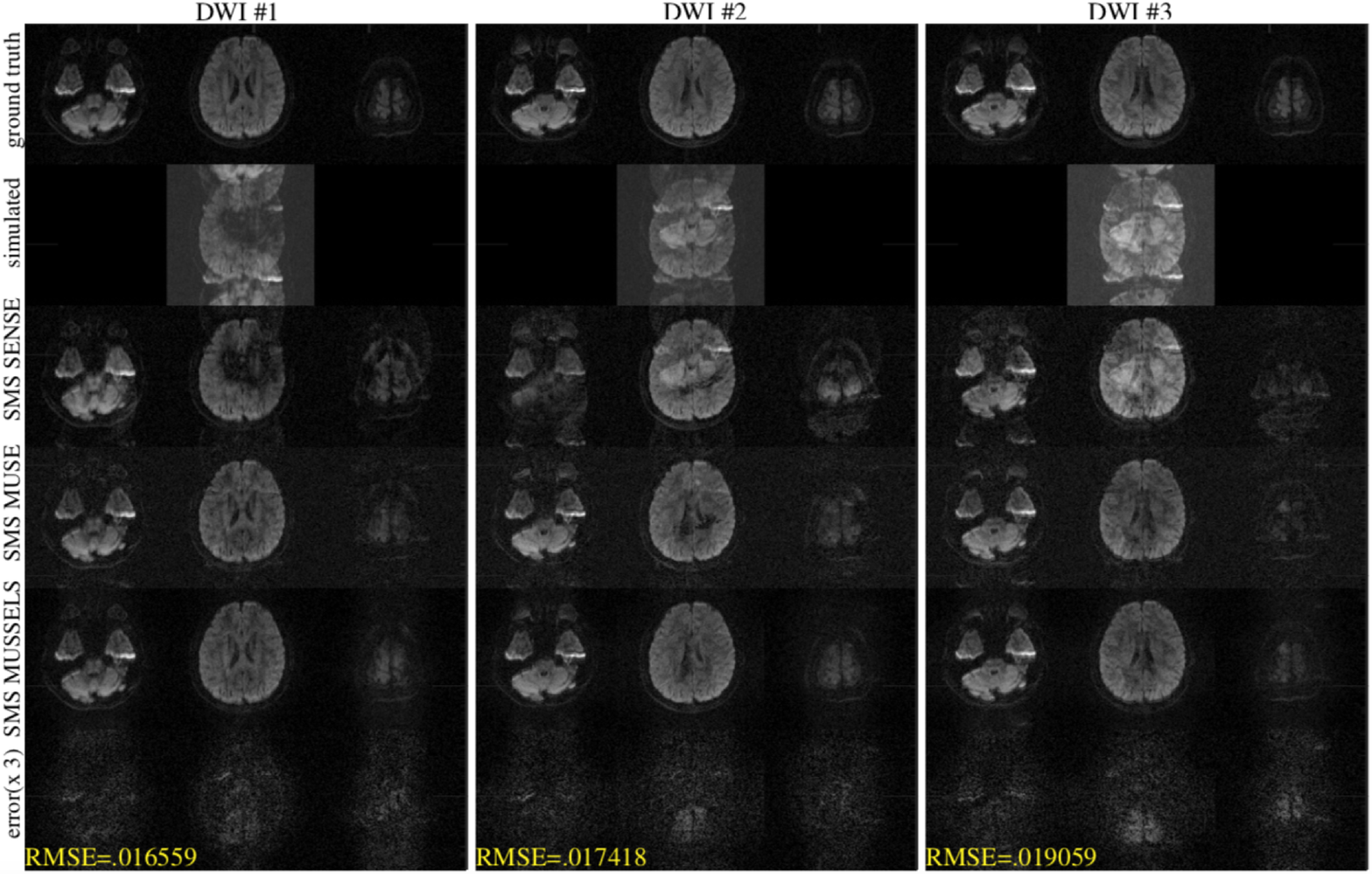}
\caption{Simulation for MB=3 from 2-shot DW data for three diffusion directions. The top row shows the three slices that were used for simulating the multi-band data. The second row shows the simulated data. The third row shows a SENSE reconstruction combing the k-space data from all shots without phase compensation. The fourth row shows the SMS MUSE reconstruction. The fifth row shows the SMS MUSSELS reconstruction and the last row shows the error in the SMS MUSSELS reconstruction compared to the ground truth.}
\label{fig:sim_ms2_mb3}
\end{figure}

\vspace{-1.5em}\noindent The next experiment studied the performance of the reconstructions at MB=3 for a 2-shot data for three different diffusion directions. The MUSSELS reconstructed images of the three slices used to simulate the multi-band data are shown in the top row of figure \ref{fig:sim_ms2_mb3}.  The k-space data of the above slices were used to simulate the multi-band data and the resulting images are shown in the second row. The SMS SENSE reconstruction (third row) show inter-shot phase errors and slice unfolding errors. While SMS MUSE was able to correct for the phase errors in the first DWI, it shows reconstruction errors for the second and third DWI. SMS MUSSELS was able to eliminate phase errors in all the images reasonably well. The difference image (magnified 3$\times$) show that some residual slice aliasing remains. However, the RMSE of the above setting is  $<$ 2\% in the worst case.\\

\begin{figure}
\includegraphics[trim = 10mm 15mm 0mm 0mm, clip, width=1\textwidth]{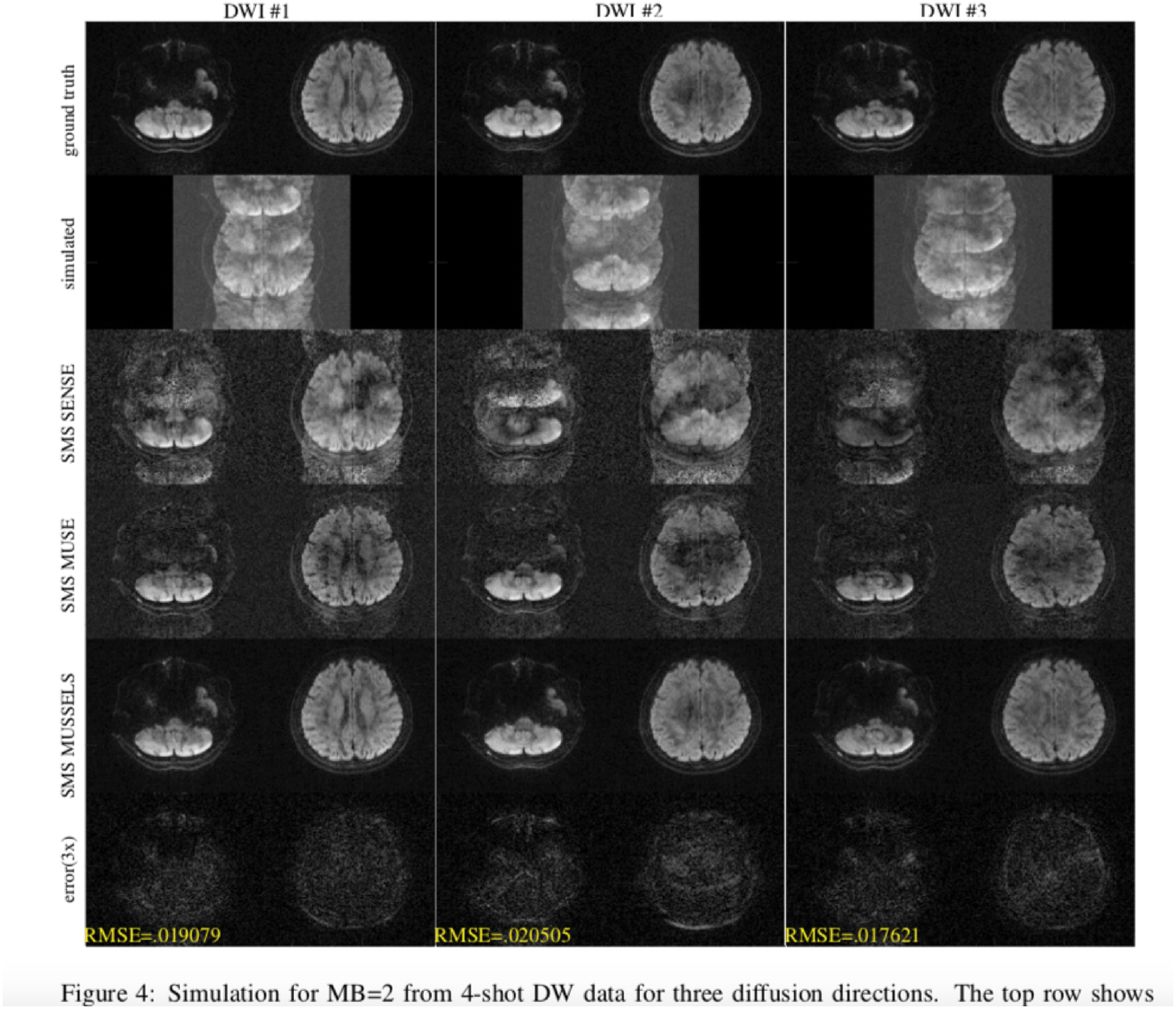}
\caption{Simulation for MB=2 from 4-shot DW data for three diffusion directions. The top row shows the two slices that were used for simulating the multi-band data. The second row shows the simulated data. The third row shows a SENSE reconstruction combing the k-space data from all shots without phase compensation. The fourth row shows the SMS MUSE reconstruction. The fifth row shows the SMS MUSSELS reconstruction and the last row shows the error in the SMS MUSSELS reconstruction compared to the ground truth.}
\label{fig:sim_ms4_mb2}
\end{figure}

\begin{figure}
\includegraphics[trim = 0mm 0mm 0mm 0mm, clip, width=1\textwidth]{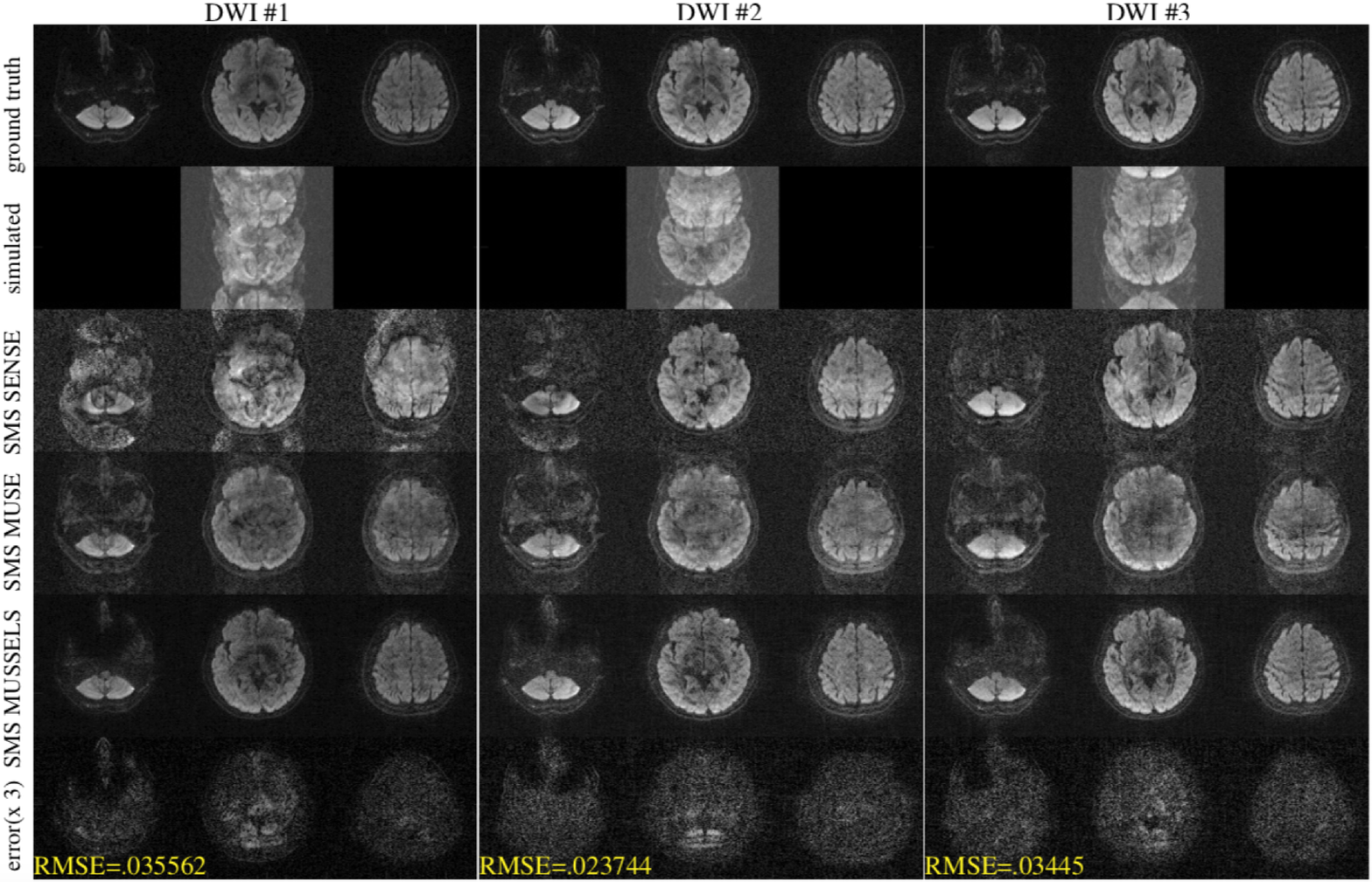}
\caption{Simulation for MB=3 from 4-shot DW data for three diffusion directions. The top row shows the three slices that were used for simulating the multi-band data. The second row shows the simulated data. The third row shows a SENSE reconstruction combing the k-space data from all shots without phase compensation. The fourth row shows the SMS MUSE reconstruction. The fifth row shows the SMS MUSSELS reconstruction and the last row shows the error in the SMS MUSSELS reconstruction compared to the ground truth.}
\label{fig:sim_ms4_mb3}
\end{figure}

\vspace{-1.5em}\noindent Similarly, we also studied the performance of the reconstructions at MB=2 and 3 using a 4-shot DW data. The various results are shown in figures \ref{fig:sim_ms4_mb2}-\ref{fig:sim_ms4_mb3}. For the case of MB=2, simulated using the 4-shot data, the SMS SENSE reconstruction show significant phase errors and slice aliasing errors. While the SMS MUSE provides improved reconstruction, residual artifacts are clearly visible in all DWIs. SMS MUSSELS reconstruction on the other hand provides significantly improved results. The error images signifies that the slice aliasing errors are minimal and phase artifacts are also not evident in the reconstructed images. The average RMSE is $\sim$ 2\% in this case. For the case of MB=3, the SMS MUSE reconstruction show significant residual artifacts. The SMS MUSSELS reconstruction can still recover the DWIs reasonably well. For some DWIs, the residual slice aliasing artifacts is present as evident from the error images. The RMSE has also increased to  $\sim$ 3.5\% on average.\\

\vspace{-1.5em}\noindent { \bf  In-vivo data}\\

\begin{figure}
\begin{minipage}[l]{0.19\textwidth}
\caption{In-vivo multi-band multi-shot data for MB=2 and Ns=2. (a) shows the images from different slice location from the multi-band acquisition before slice unfolding and phase correction. (b) shows a SMS SENSE reconstruction,  (c) shows the SMS MUSE reconstruction and (d) shows the SMS MUSSELS reconstruction.}
\label{fig:ms2_mb2}
\end{minipage}
\begin{minipage}[r]{0.79\textwidth}
\includegraphics[trim = 0mm 0mm 0mm 0mm, clip, width=1\textwidth]{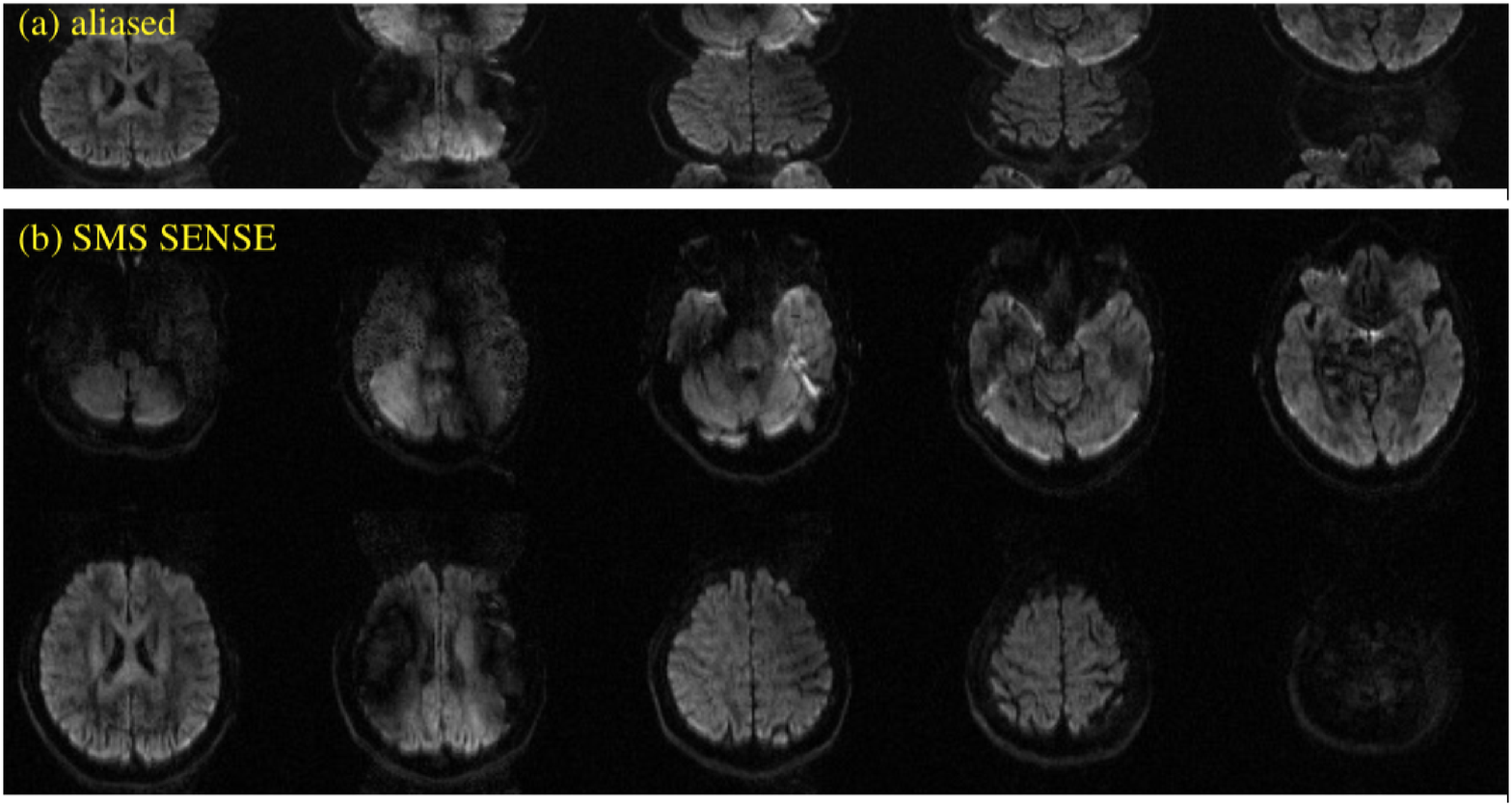}
\includegraphics[trim = 0mm 0mm 0mm 0mm, clip, width=1\textwidth]{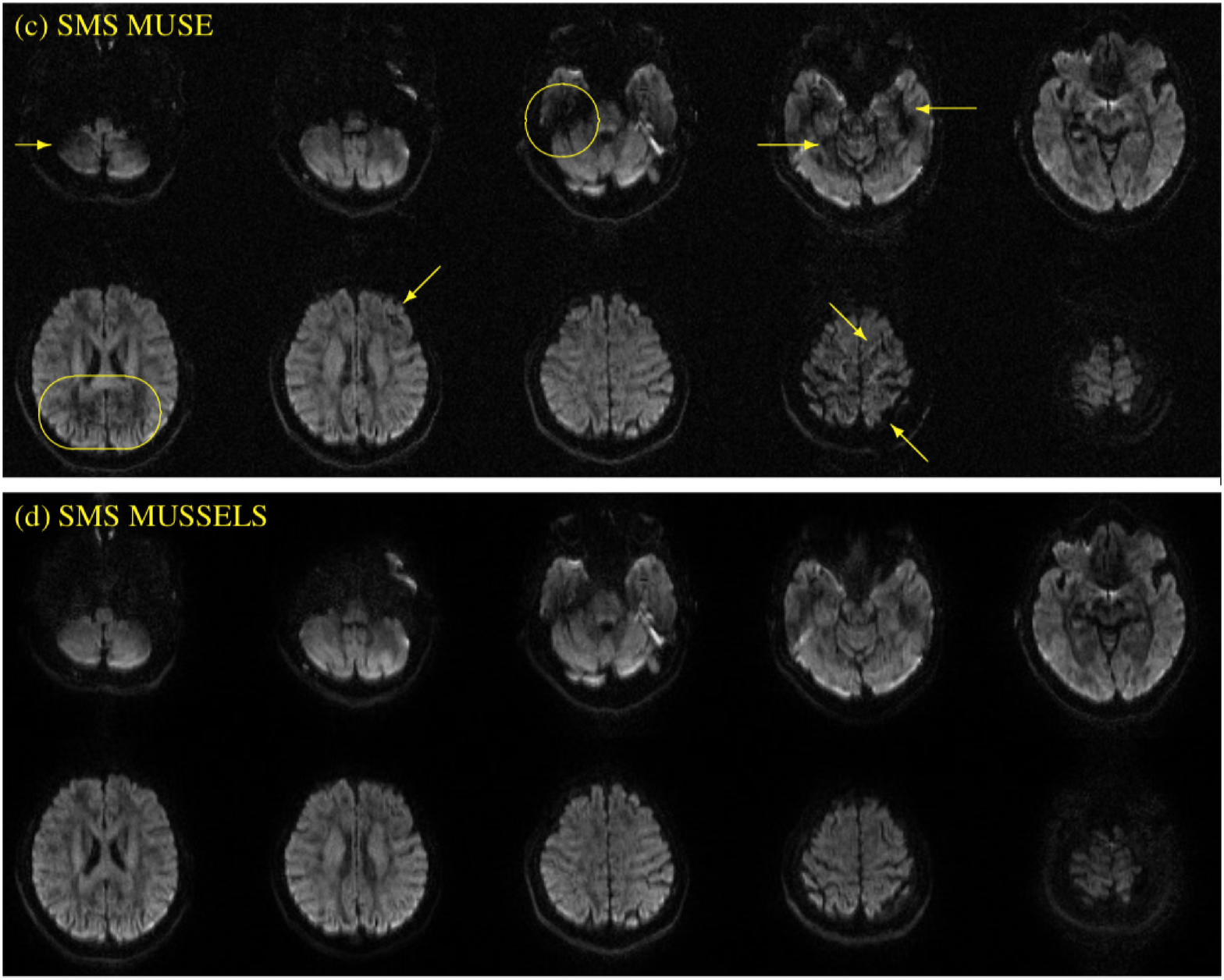}
\end{minipage}
\end{figure}

\vspace{-1.5em}\noindent  Figure \ref{fig:ms2_mb2} shows the various reconstructions performed on multi-band in-vivo data. This data was collected at MB~=~2 using a 2-shot readout. The top row shows the SOS reconstruction of the inverse Fourier transformed k-space data acquired using the modified blipped CAIPI sequence. Images from a subset of the sampled slice locations are shown for a selected diffusion direction. The SMS SENSE reconstruction show  artifacts in various slice locations. The SMS MUSE reconstruction was able to recover the images reasonably well; however the image quality is suboptimal (highlighted by the arrows) in some regions. The SMS MUSSELS reconstruction show robust reconstruction in those regions as well in all slice locations. For generalizability of the above results, reconstruction from a few other diffusion directions are given in the supplementary data (Fig S1, Fig S2).\\

\begin{figure}
\begin{minipage}[l]{0.4985\textwidth}
\vspace{28mm}
\caption{In-vivo multi-band multi-shot data for MB=3 and Ns=2. (a) shows the images from different slice location from the multi-band acquisition before slice unfolding and phase correction. (b) shows a SMS SENSE reconstruction,  (c) shows the SMS MUSE reconstruction and (d) shows the SMS MUSSELS reconstruction.}
\label{fig:ms2_mb3}
\includegraphics[trim = 0mm 0mm 290mm 220mm, clip, width=1\textwidth]{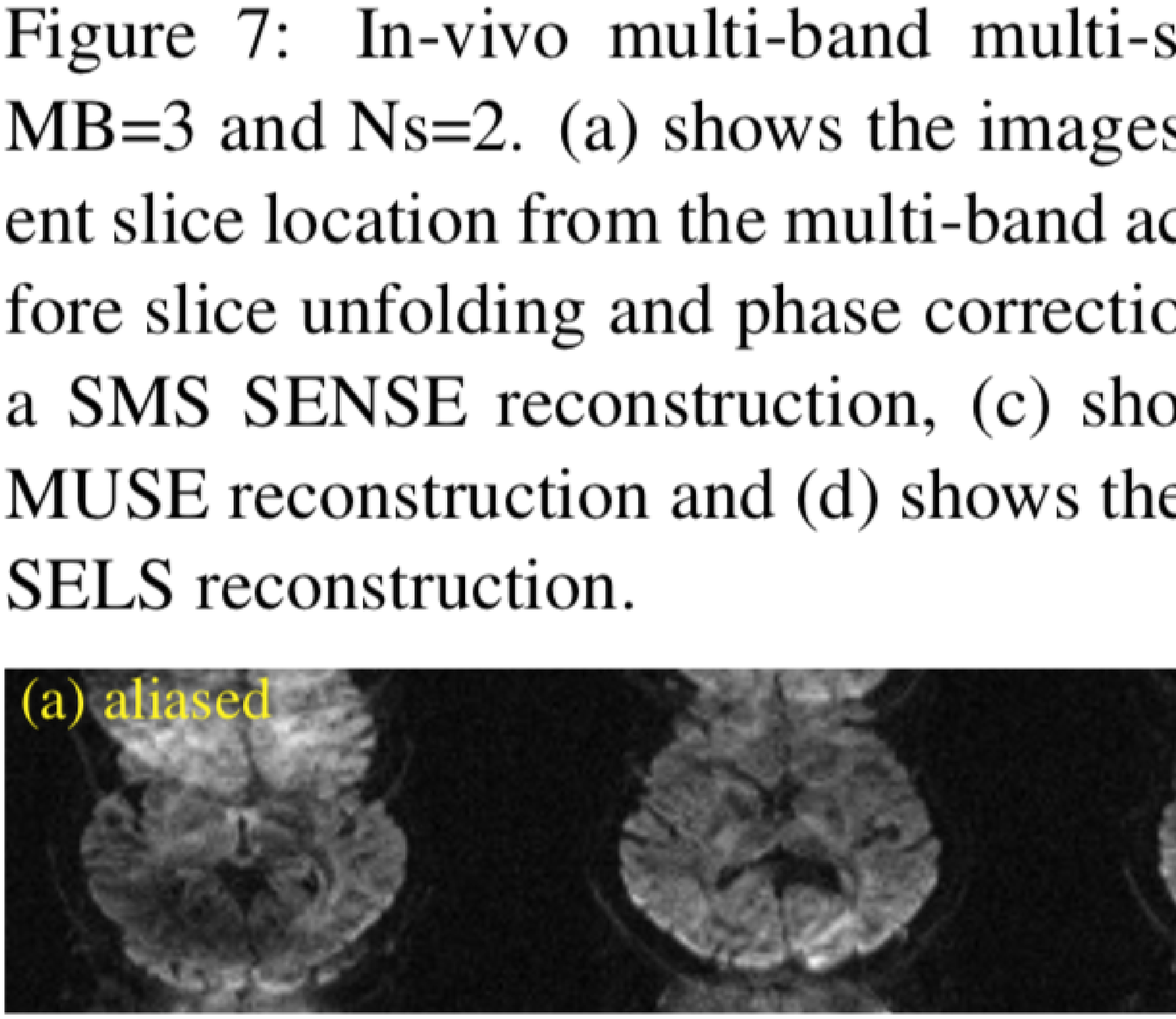}
\end{minipage}
\begin{minipage}[r]{0.4985\textwidth}
\includegraphics[trim = 290mm 0mm 0mm 0mm, clip, width=1\textwidth]{fig7a.eps}
\end{minipage}
\includegraphics[trim = 0mm 0mm 0mm 0mm, clip, width=1\textwidth]{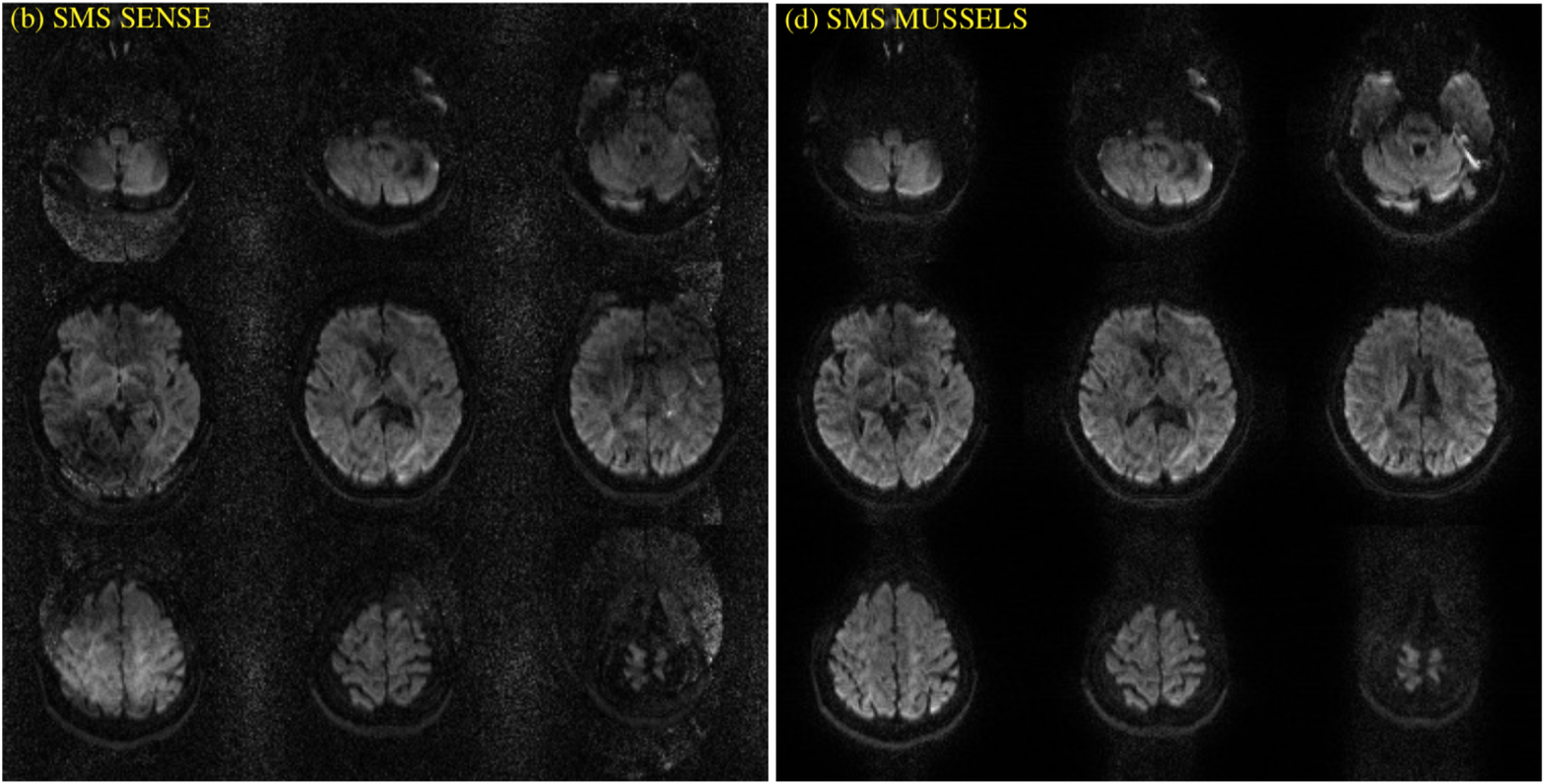}
\end{figure}

\vspace{-1.5em}\noindent  At MB=3, the superior performance of SMS MUSSELS is clearly visible. Figure \ref{fig:ms2_mb3} shows the various reconstructions performed on in vivo data collected at a multi-band factor of 3 using a 2-shot read-out. Residual artifacts are clearly visible on the SMS MUSE reconstructions. SMS MUSSELS method provides robust reconstructions at all slice locations and different diffusion directions (please see supplementary data, Fig S3, Fig S4)).\\

\begin{figure}[b!]
\begin{minipage}[l]{0.4985\textwidth}
\vspace{28mm}
\caption{In-vivo multi-band multi-shot data for MB=3 and Ns=4. (a) shows the images from different slice location from the multi-band acquisition before slice unfolding and phase correction. (b) shows a SMS SENSE reconstruction,  (c) shows the SMS MUSE reconstruction and (d) shows the SMS MUSSELS reconstruction.}
\label{fig:ms4_mb3}
\includegraphics[trim = 0mm 0mm 290mm 220mm, clip, width=1\textwidth]{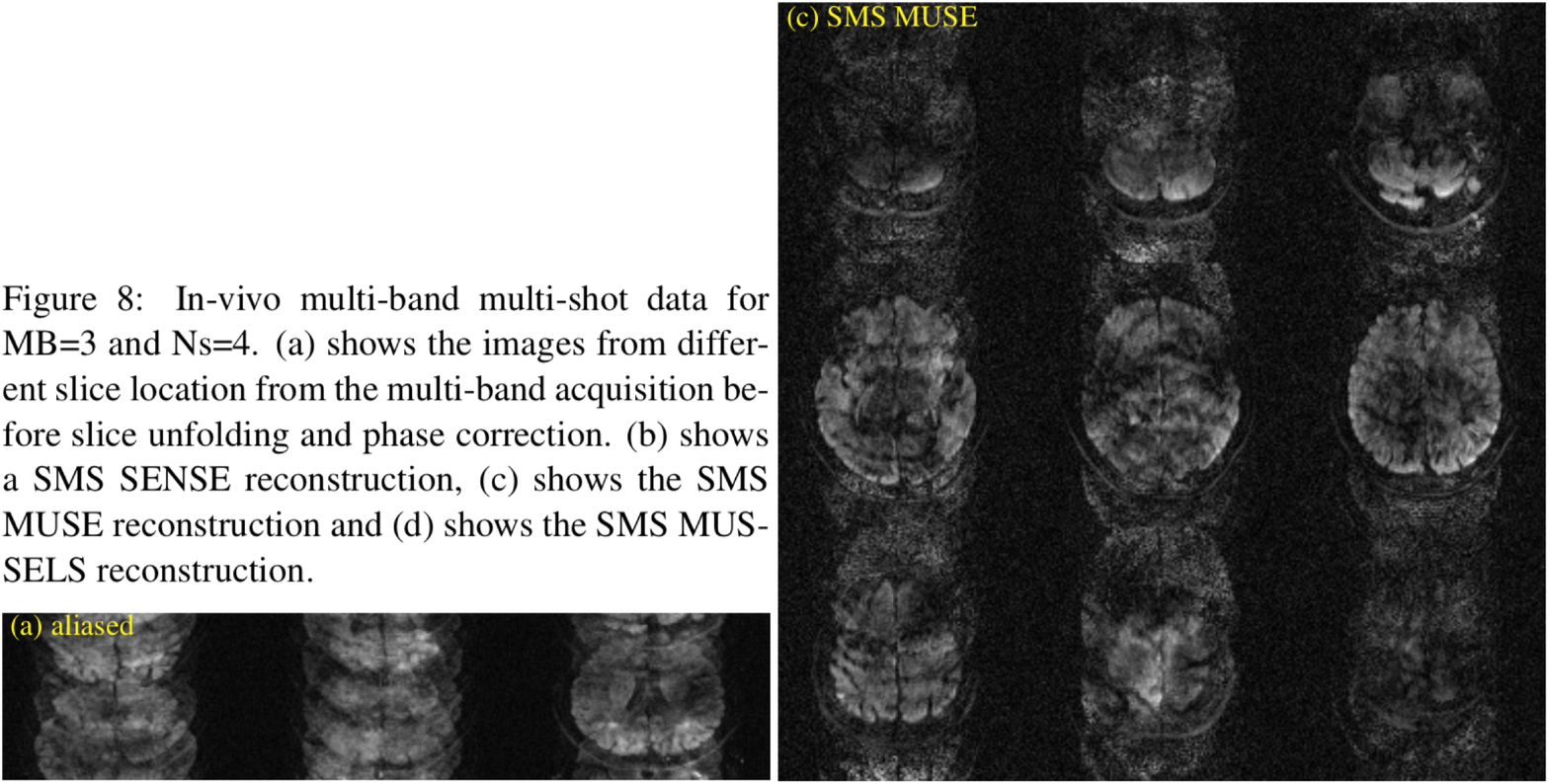}
\end{minipage}
\begin{minipage}[r]{0.4985\textwidth}
\includegraphics[trim = 290mm 0mm 0mm 0mm, clip, width=1\textwidth]{fig8a.eps}
\end{minipage}
\includegraphics[trim = 0mm 0mm 0mm 0mm, clip, width=1\textwidth]{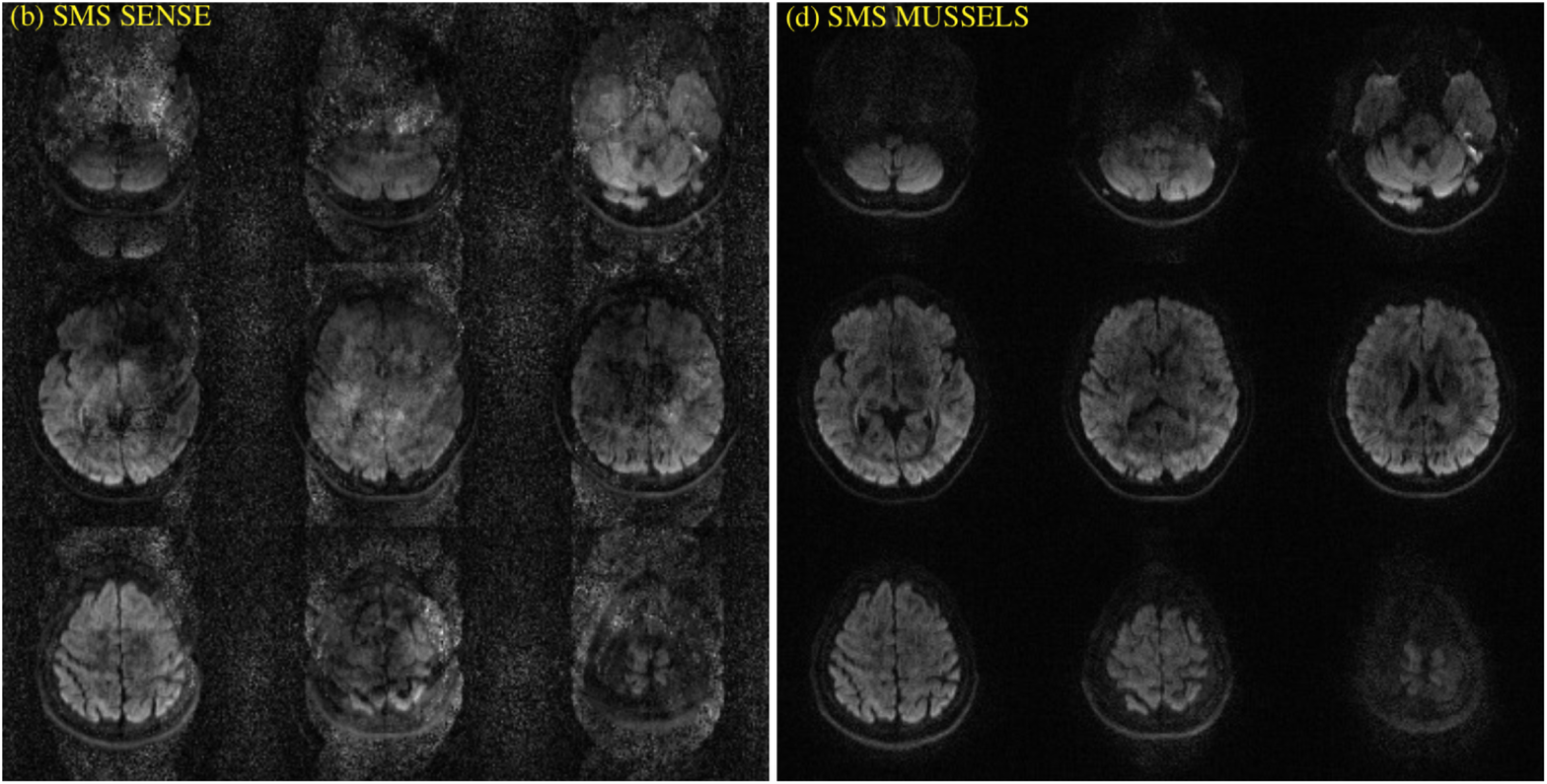}
\end{figure}

\begin{figure}[b!]
\includegraphics[trim = 0mm 0mm 0mm 0mm, clip, width=1\textwidth]{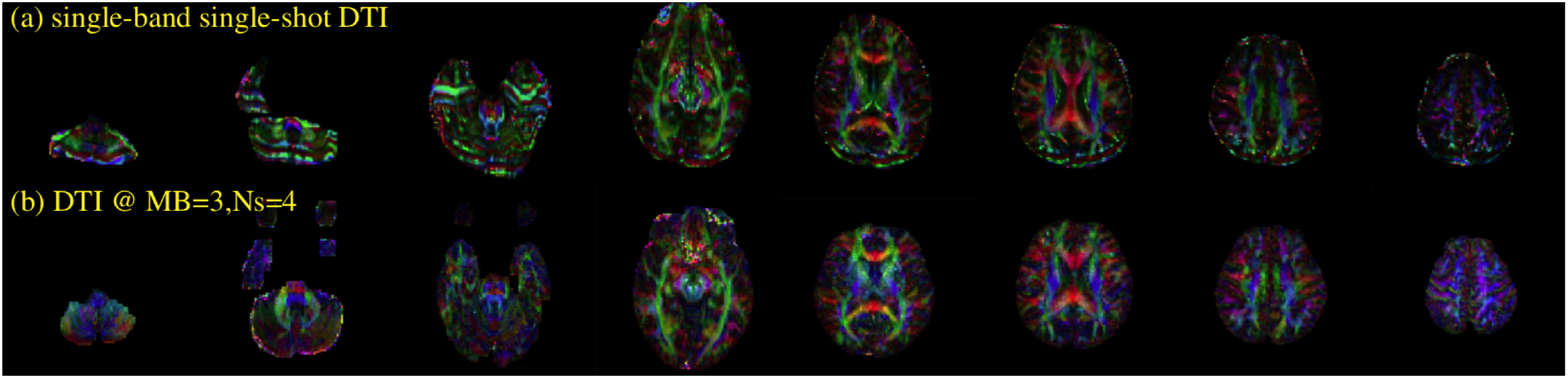}
\caption{The fiber direction color encoded maps from a tensor fitting for the cases of (a) single-band single-shot data and (b) multi-band multi-shot data. Owing to separate acquisitions, there is a slight mismatch of slice encodings. However, the geometric distortions effects and the image quality show significant improvement for the latter case especially for the inferior slices.}
\label{fig:FA}
\end{figure}

\vspace{-1.5em}\noindent  Finally, we also provide various reconstructions on an in-vivo 4-shot DW data that was collected at a multi-band factor of 3.  Figure \ref{fig:ms4_mb3} shows the various reconstructions. At high multi-shot and multi-band factors, there is substantial residual artifacts in the SMS MUSE reconstructions. SMS MUSSELS algorithm was much more successful in mitigating the errors at this challenging case. Additional results from other diffusion directions are provided in supplementary data (Fig S5, Fig S6). A diffusion tensor was fit to the DWIs reconstructed using the proposed method to compute the fractional anisotropy maps and the direction encoded color maps. The resulting images are also provided for this data in figure \ref{fig:FA}. For comparison, the fiber direction encoded color maps from a traditional single-band single shot acquisition is also provided. As expected, the single-shot data show severe artifacts at the inferior slices and geometric distortion in all slices, which are mitigated to a great extend in the multi-shot data.\\

\vspace{-0.5em}\noindent {\Large Discussion}\\  

\vspace{-1.5em}\noindent Although ms-DWI is a better alternative to single shot DWI for high resolution diffusion imaging, the long scan time associated with ms-DWI makes it hard to be used in studies requiring high angular resolution. To improve the time-efficiency of ms-DWI, simultaneous multi slice imaging had been proposed. However, the associated problem of recovering DWIs that are devoid of phase and slice aliasing artifact is challenging. In this work, we presented a novel reconstruction method to address the above reconstruction problem. Motivated by our previous work on ms-DWI reconstruction, we developed a new algorithm that exploited the low-rank property of multi-shot data to simultaneously unfold the slices and correct for phase related artifacts in a joint scheme. We demonstrate the superior performance of the proposed algorithm in recovering SMS-accelerated ms-DW data both using simulations and in-vivo experiments.\\

\vspace{-1.5em}\noindent In the absence of a ground truth data for studying ms-DW experiments, we relied on simulations to study the performance of the proposed reconstructions. We showed that the SMS MUSSELS reconstruction could recover challenging cases of 4-shot DW data at a multi-band factor of 3 reasonably well with an RMSE less than 4\%. The above reconstruction was performed without the use of phase navigators or phase calibration. This was made possible by joint reconstruction framework that incorporates (i) SMS slice unfolding via a forward model that performs spatial slice shifting, and (ii) joint k-space data recovery using a structured low-rank matrix completion. Our experiments using in-vivo data show very similar performance to the simulation studies. Compared to SMS MUSSELS, the SMS MUSE method for simultaneous slice unfolding and phase correction is ill-conditioned and show significant noise amplification as the multi-band and multi-shot factor increases. However, the additional reconstruction penalty in the SMS MUSSELS that enforces low-rankedness of a block Hankel matrix independently for each slice provided a robust reconstruction scheme for such cases. The low-rankedness proves the existence of shift-invariant null-space filters in the frequency domain that when convolve with the sampled data points annihilate the signals. The low-rankedness of the {\it fully known} block-Hankel matrix, thus, provides a powerful reconstruction constraint to recover missing k-space samples in each of the multi-shot k-space acquisitions. Moreover, the above image recovery constraint can be employed to recover missing data in several situations where the acquired images only differ in their phase. This property has been exploited in several image reconstruction problems including image ghost correction \cite{Mani2017,Mani2016d}, correction of trajectory errors \cite{Mani2018} and correction of B0 inhomogeneity \cite{Balachandrasekaran2018}.  \\

\vspace{-1.5em}\noindent Our simulations and in-vivo study show that as the multi-band factor increases, it becomes increasingly hard to unfold the signals. Notably it is not the inter-shot phase that is causing the problem, rather the SMS acquisition itself since the b0 images also exhibit signal loss in the center slices (images not shown).  Several factors can contribute to this. It is likely that the coil sensitivity distribution across the axial slices may not be heterogeneous over a slice gap of 45mm to effectively unalias the slices. Others have suggested acquiring data sagittally to better use the variability in the coil sensitivity distribution in that orientation \cite{Frost2015,Dai2017}. In the current work, we relied on coil sensitivity estimates that were obtained from b0 scans in all the reconstructions since the distortions were matched this way and this approach required no additional calibration scans. Better coil sensitivity estimates with improved SNR from separate calibration scans may improve the reconstruction results. Another reason could be related to the g-factor of multi-shot acquisition. Since there is additional aliasing of the different shots in addition to the intentional slice aliasing, the same performance obtained in a single-shot acquisition may not translate to multi-shot acquisition. Similar observation has been made for in-plane accelerated single-shot multi-band imaging also \cite{Setsompop2012,Feinberg2013,Barth2016}. Notably, the proposed SMS MUSSELS method can be modified to recover other spatial shifting factors also, by modifying the phase factor appropriately in the forward model. \\

\vspace{-1.5em}\noindent We employed partial Fourier acquisition in all of our data acquisitions to achieve the minimum possible TE. To recover missing data in such scenario, we had previously introduced additional regularizations in the MUSSELS framework. The same regularization can be extended to the SMS MUSSELS framework also and is expected to improve the image quality of the reconstructions. However, we did not pursue this approach in the current work due to the long reconstruction times involved. However, several enhancements are possible to speed up the general MUSSELS formulations which we will be pursing in our future work. Of note, a fast Hankel matrix based SMS slice unfolding has been proposed earlier \cite{Park2017}.  Similar methods can be extended for multi-shot diffusion weighted data also using faster implementations.  \\

\vspace{-1.5em}\noindent  Adopting SMS acceleration for multi-shot imaging brings down the time to perform a 4-shot acquisition (3.58 mins) comparable to that of a traditional single-shot single-band scan (3.23 mins) for our experiments. This is accompanied by drastic improvement in image quality as observed from Figure 9.  Thus, SMS-accelerated ms-DWI can provide a time-efficient way to extend ms-DWI scans for high angular resolution diffusion imaging studies. The proposed reconstruction is shown to enable such studies for routine imaging applications. \\

\vspace{-1.5em}\noindent In conclusion, we proposed a new reconstruction method to simultaneously slice unfold and jointly recover multi-shot diffusion data using a navigator-free approach. We demonstrated the ability of SMS MUSSELS to effectively reconstruct 4-shot DW data accelerated using a multi-band factor of 3. Such acceleration factors can improve the time efficiency of ms-DW acquisitions to enable their utility in high angular resolution diffusion studies. The proposed reconstruction is generalizable to non-Cartesian trajectories also.\\

\vspace{2mm}
{\bf{{Acknowledgements}}}
\vspace{2mm}

Financial support for this study was provided by grants NIH 5 R01 EB022019, 5 R01 MH111578 and NIH 1R01EB019961-01A1.

\newpage
\clearpage

\newpage
\clearpage

\newpage

\newpage

\newpage
\clearpage

\newpage

\newpage
\clearpage

\newpage
\clearpage

\newpage
\clearpage

\newpage
{\bf{Legends: }}\\
Fig 1: Forward model for SMS MUSSELS involves taking (i) the inverse Fourier transform (${\cal {F}}^{-1}$) of the k-space data of L slices, (ii) multiplying the resulting image data by the coil sensitivities (${\cal {S}}$), (iii) taking the Fourier transform (${\cal {F}}$) to get the k-space channel-by-channel data of each slice, (iv) introducing slice shift by multiplying each slice data by phase factors ($\theta_l$) and adding the data from all slices in a channel-by-channel manner and finally (v) multiplying each channel data by a sampling mask (${\cal {M}}$) corresponding to the multi-shot acquisition.\\

Fig 2: Simulation for MB=2 from 2-shot DW data for three diffusion directions. The top row shows the two slices that were used for simulating the multi-band data. The second row shows the simulated data. The third row shows a SENSE reconstruction combing the k-space data from all shots without phase compensation. The fourth row shows the SMS MUSE reconstruction. The fifth row shows the SMS MUSSELS reconstruction and the last row shows the error in the SMS MUSSELS reconstruction compared to the ground truth.\\

Fig 3: Simulation for MB=3 from 2-shot DW data for three diffusion directions. The top row shows the three slices that were used for simulating the multi-band data. The second row shows the simulated data. The third row shows a SENSE reconstruction combing the k-space data from all shots without phase compensation. The fourth row shows the SMS MUSE reconstruction. The fifth row shows the SMS MUSSELS reconstruction and the last row shows the error in the SMS MUSSELS reconstruction compared to the ground truth.\\

Fig 4: Simulation for MB=2 from 4-shot DW data for three diffusion directions. The top row shows the two slices that were used for simulating the multi-band data. The second row shows the simulated data. The third row shows a SENSE reconstruction combing the k-space data from all shots without phase compensation. The fourth row shows the SMS MUSE reconstruction. The fifth row shows the SMS MUSSELS reconstruction and the last row shows the error in the SMS MUSSELS reconstruction compared to the ground truth.\\

Fig 5: Simulation for MB=3 from 4-shot DW data for three diffusion directions. The top row shows the three slices that were used for simulating the multi-band data. The second row shows the simulated data. The third row shows a SENSE reconstruction combing the k-space data from all shots without phase compensation. The fourth row shows the SMS MUSE reconstruction. The fifth row shows the SMS MUSSELS reconstruction and the last row shows the error in the SMS MUSSELS reconstruction compared to the ground truth.\\

Fig 6: In-vivo multi-band multi-shot data for MB=2 and Ns=2. (a) shows the images from different slice location from the multi-band acquisition before slice unfolding and phase correction. (b) shows a SMS SENSE reconstruction,  (c) shows the SMS MUSE reconstruction and (d) shows the SMS MUSSELS reconstruction.\\

Fig 7: In-vivo multi-band multi-shot data for MB=3 and Ns=2. (a) shows the images from different slice location from the multi-band acquisition before slice unfolding and phase correction. (b) shows a SMS SENSE reconstruction,  (c) shows the SMS MUSE reconstruction and (d) shows the SMS MUSSELS reconstruction.\\

Fig 8: In-vivo multi-band multi-shot data for MB=3 and Ns=4. (a) shows the images from different slice location from the multi-band acquisition before slice unfolding and phase correction. (b) shows a SMS SENSE reconstruction,  (c) shows the SMS MUSE reconstruction and (d) shows the SMS MUSSELS reconstruction.\\

Fig 9: The fiber direction color encoded maps from a tensor fitting for the cases of (a) single-band single-shot data and (b) multi-band multi-shot data. Owing to separate acquisitions, there is a slight mismatch of slice encodings. However, the geometric distortions effects and the image quality show significant improvement for the latter case especially for the inferior slices..\\

Fig S1: In-vivo multi-band multi-shot data for MB=2 and Ns=2 from other diffusion directions. (a) shows the images from different slice location from the multi-band acquisition before slice unfolding and phase correction. (b) shows a SMS SENSE reconstruction,  (c) shows the SMS MUSE reconstruction and (d) shows the SMS MUSSELS reconstruction.\\

Fig S2: In-vivo multi-band multi-shot data for MB=2 and Ns=2 from other diffusion directions. (a) shows the images from different slice location from the multi-band acquisition before slice unfolding and phase correction. (b) shows a SMS SENSE reconstruction,  (c) shows the SMS MUSE reconstruction and (d) shows the SMS MUSSELS reconstruction.\\

Fig S3: In-vivo multi-band multi-shot data for MB=3 and Ns=2 from other diffusion directions. (a) shows the images from different slice location from the multi-band acquisition before slice unfolding and phase correction. (b) shows a SMS SENSE reconstruction,  (c) shows the SMS MUSE reconstruction and (d) shows the SMS MUSSELS reconstruction.\\

Fig S4: In-vivo multi-band multi-shot data for MB=3 and Ns=2 from other diffusion directions. (a) shows the images from different slice location from the multi-band acquisition before slice unfolding and phase correction. (b) shows a SMS SENSE reconstruction,  (c) shows the SMS MUSE reconstruction and (d) shows the SMS MUSSELS reconstruction.\\

Fig S5: In-vivo multi-band multi-shot data for MB=3 and Ns=4 from other diffusion directions. (a) shows the images from different slice location from the multi-band acquisition before slice unfolding and phase correction. (b) shows a SMS SENSE reconstruction,  (c) shows the SMS MUSE reconstruction and (d) shows the SMS MUSSELS reconstruction.\\

Fig S6: In-vivo multi-band multi-shot data for MB=3 and Ns=4 from other diffusion directions. (a) shows the images from different slice location from the multi-band acquisition before slice unfolding and phase correction. (b) shows a SMS SENSE reconstruction,  (c) shows the SMS MUSE reconstruction and (d) shows the SMS MUSSELS reconstruction.\\

\newpage
{\small
\bibliographystyle{unsrtnat} 
\bibliography{sms_mussels}}

\end{document}